\documentclass[shortnote,twocolumn]{jpsj3}
\usepackage{txfonts}

\usepackage{color}

\usepackage{ulem}
\usepackage{siunitx}
\usepackage{comment}
\usepackage{ulem}
\usepackage{xcolor}
\DeclareRobustCommand{\erase}{\bgroup\markoverwith{\textcolor{blue}{\rule[.5ex]{2pt}{0.8pt}}}\ULon}%{0.4pt}

\title{Emergence of Chiral Helimagnetic Order in Chromium-intercalated Tantalum Disulfide CrTa$_{3}$S$_{6}$ Powders with Controlled Intercalation}

\author{
Ayaka Toshima$^{1}$,
Keigo Mizutani$^{1}$,
Yusuke Kousaka$^{1}$\thanks{koyu@omu.ac.jp},
Kazuki Ohishi$^{2}$,
Hiroaki Shishido$^{3}$,
Yoshihiko Togawa$^{1}$
}
\inst{
$^{1}$ Department of Physics and Electronics, Osaka Metropolitan University, Sakai, Osaka 599-8531, Japan \\
$^{2}$ Neutron Science and Technology Center, Comprehensive Research Organization for Science and Society (CROSS), Tokai, Ibaraki 319-1106, Japan \\
$^{3}$ Center for Shared Research Equipment, Osaka Metropolitan University, Sakai, Osaka 599-8531, Japan %\\
}

\abst{
We report a highly sensitive change in magnetic properties of a chiral Cr-intercalated transition-metal dichalcogenide Cr$_{x}$Ta$_{3}$S$_{6}$.
Magnetization curves and small-angle neutron scattering data revealed that the powder samples exhibit chiral helimagnetism with a Cr intercalation quantity $x$ below 0.996, while they show ferromagnetism above 1.000.
The emergence and temperature-dependent evolution of the helimagnetic period are argued in terms of sample dimensions of powders and microfabricated crystals.
}

\begin{document}
\maketitle

Chiral helimagnets have attracted much attention due to a formation of 
chiral spin-modulated structures such as chiral soliton lattice (CSL)\cite{Kishine2005, Togawa2012} and chiral vortices (magnetic Skyrmions)\cite{Bogdanov1994, Muhlbauer2009}.
In most cases, they are derived from a chiral helical magnetic order (CHM) in the presence of a magnetic field. Thus, the CHM detection is an important issue to deepen the understanding of physics behind chiral helimagnetism.
The CHM period is determined by a ratio of symmetric Heisenberg and antisymmetric Dzyaloshinskii-Moriya (DM) exchange interactions\cite{Dzyaloshinskii1958, Moriya1960} and inevitably results in a long periodicity, being tens of nanometers in typical chiral helimagnets.
In this respect, small-angle neutron scattering (SANS) is one of the powerful experimental methods for detecting a long-range nature of the CHM, applicable to both single crystalline and polycrystalline samples.

\begin{figure}[t]
\centering
\includegraphics[width=7.4cm]{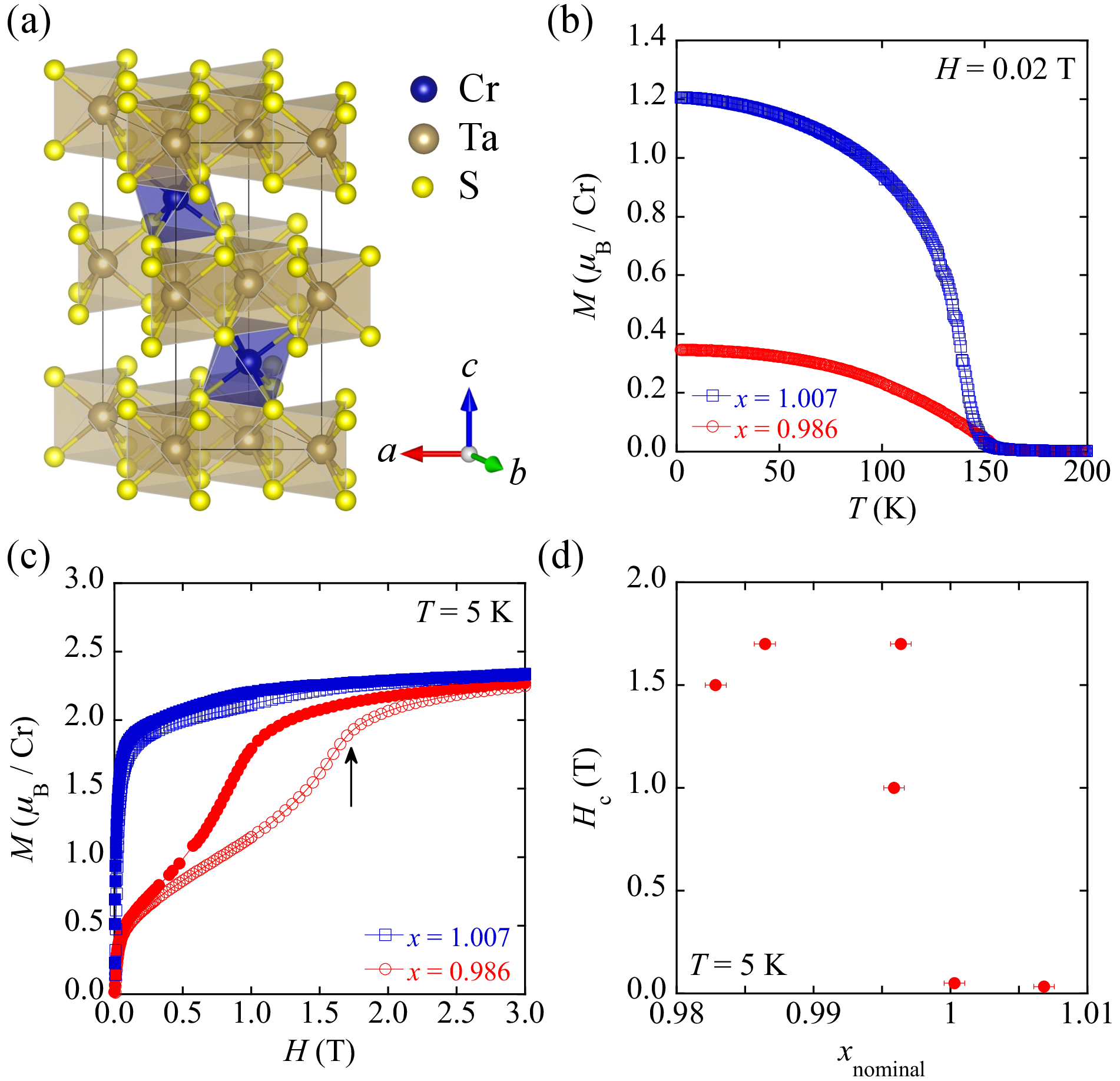}
\caption{
(Color online) (a) Crystal structure of CrTa$_{3}$S$_{6}$. (b) Temperature dependence of magnetization at \SI{0.02}{\tesla} in the polycrystalline Cr$_{x}$Ta$_{3}$S$_{6}$ samples with $x = 0.986$ and $1.007$. (c) Magnetization curves at \SI{5}{\kelvin} in the $H$-increase and decrease processes, which are represented by open and closed marks, respectively. 
The vertical arrow indicates the critical magnetic field $H_{\rm c}$ in the $H$-increase process at $x = 0.986$. (d) $H_{\rm c}$ as a function of the Cr amount $x$.
}
\label{f1}
\end{figure}

A Cr-intercalated transition-metal dichalcogenide Cr$M_{3}$S$_{6}$ ($M$: Nb or Ta) is one of the representative chiral helimagnets which forms a chiral crystal structure, as shown in Fig.~\ref{f1}(a), and can host the CHM and CSL.
Importantly, magnetic properties of the Cr$M_{3}$S$_{6}$ single crystals are sensitive to the sample quality, in particular to an amount of Cr intercalation.
In the case of CrNb$_{3}$S$_{6}$, its deviation from the ideal value by a few percent favors the ferromagnetism over the CHM and shifts the value of a critical temperature $T_{\rm c}$ by several tens of kelvins\cite{Kousaka2022}.
A similar influence should occur in CrTa$_{3}$S$_{6}$, where indeed the reported values of a critical magnetic field $H_{\rm c}$ vary from 1.4 to \SI{1.7}{\tesla} in the single crystals\cite{Zhang2021, Obeysekera2021, Mizutani2023}.
Such a sample-dependent discrepancy remains unresolved and highlights the challenge of synthesizing high-quality CrTa$_{3}$S$_{6}$ crystals.

In this short note, we investigated the influence of Cr intercalation quantity $x$ on the magnetic order of polycrystalline Cr$_{x}$Ta$_{3}$S$_{6}$ based on magnetization and SANS data.
A slight difference in $x$ resulted in drastic changes in $H_{\rm c}$ and stabilization of CHM and CSL instead of ferromagnetism.

Polycrystalline powder samples of Cr$_{x}$Ta$_{3}$S$_{6}$ were synthesized by a gas phase method.
Powders of Cr, Ta and S were accurately weighed in a molar ratio of $x$ : 3 : 6 and homogeneously mixed using an alumina mortar.
The mixture was sealed in an evacuated quartz tube and heated in an electric tube furnace at 1000 $^{\circ}$C for a week.
The nominal Cr amount $x$ was chosen to be slightly different from a stoichiometric value, ranging from 0.983 to 1.007.

Magnetic properties of the obtained polycrystalline samples were evaluated by magnetization measurements using a superconducting quantum interference device
magnetometer (Quantum Design MPMS-3).
The magnetization was measured as a function of temperature $T$ in the field-cooling (FC) processes. The magnetization was also examined with increasing and decreasing an external magnetic field $H$ at different $T$.
To investigate the magnetic structure, zero-field SANS experiments were performed at BL15 (Small and wide angle neutron scattering instrument TAIKAN)\cite{TAIKAN2015} at the Material and Life Science Facility (MLF) in J-PARC, Japan.
Azimuthally averaged profiles, collected by two-dimensional position sensitive detectors, were processed using Utsusemi software\cite{Inamura2013}.

Figure~\ref{f1}(b) shows the $T$ dependence of the magnetization at a fixed $H$ of \SI{0.02}{\tesla} in the Cr$_{x}$Ta$_{3}$S$_{6}$ powder samples.
The magnetization starts to grow at a critical temperature $T_{\rm c}$ of \SI{150}{\kelvin} while decreasing $T$ in all the samples examined irrespective of the Cr amount $x$.
However, it turns out that different magnetic order was formed depending on $x$, as seen in the magnetization curves at \SI{5}{\kelvin} in Fig.~\ref{f1}(c).
The sample with $x$ of 1.007 exhibited ferromagnetic behavior with the magnetization saturating at the low-$H$ region.
On the other hand, the sample with $x$ of 0.986 showed two-step changes in the magnetization:
a ferromagnetic increase was observed in the low-$H$ region, while a downward-convex increase appeared in the high-$H$ region.
It saturated at \SI{1.7}{\tesla} in the $H$-increase process, as indicated by an arrow in Fig.~\ref{f1}(c).
This kink structure corresponds to $H_{\rm c}$ for the CSL formation, as observed in CrNb$_{3}$S$_{6}$ single crystals\cite{Kousaka2009}.
The obtained $H_{\rm c}$ value is consistent with that of the CrTa$_{3}$S$_{6}$ single crystals.\cite{Mizutani2023}
The magnetization curve shifts toward lower $H$ and shows a reduction in the $H_{\rm c}$ value in the $H$-decrease process, which is ascribed to the presence of a surface barrier for the soliton penetration during the CSL formation.\cite{Mizutani2023, Shinozaki2018}
Figure~\ref{f1}(d) summarizes the $H_{\rm c}$ distribution as a function of $x$.
A sharp increase in $H_{\rm c}$ occurs at $x$ less than 0.996. 
It is clear that a small difference in the Cr nominal amount affects magnetic properties in CrTa$_{3}$S$_{6}$.

\begin{figure}[t]
\centering
\includegraphics[width=7.4cm]{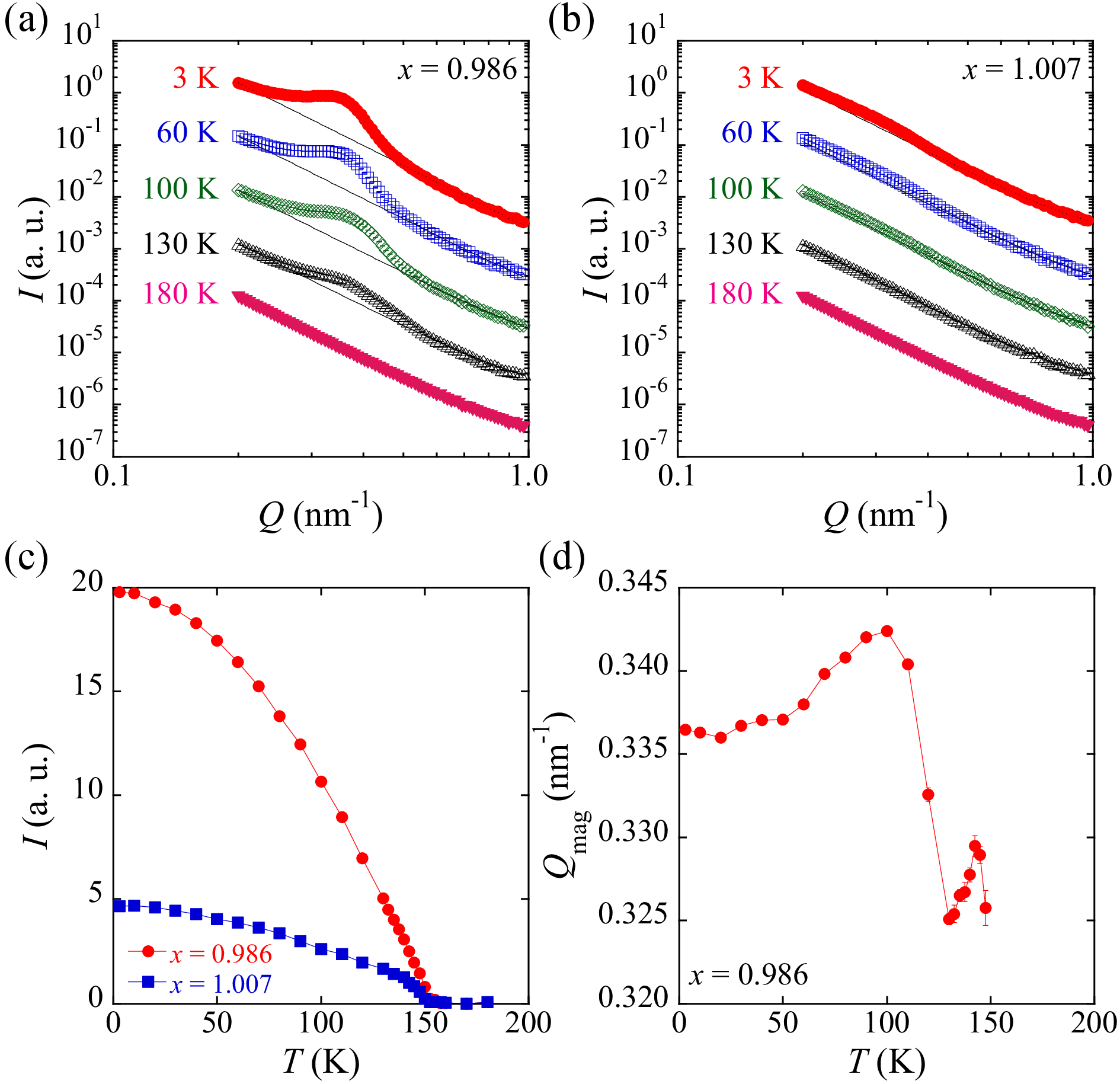}
\caption{
(Color online) SANS profiles of the polycrystalline Cr$_{x}$Ta$_{3}$S$_{6}$ at (a) $x = 0.986$ and (b) $x = 1.007$.
The solid lines represent background signals derived from Porod's law.
(c) Temperature dependence of the magnetic signals for both samples, processed by subtracting the SANS profiles in paramagnetic phase.
(d) Temperature dependence of the magnetic satellite peak position at $x = 0.986$.}
\label{f2}
\end{figure}

SANS profiles, collected for the powder samples with $x$ of 0.986 and 1.007, revealed a difference in the magnetic structures.
Wave-vector $Q-$scan profiles show a magnetic satellite peak attributed to CHM formation below $T_{\rm c}$ for the sample with $x$ of 0.986 in Fig.~\ref{f2}(a), while no magnetic satellite peak is observed for the sample with $x$ of 1.007 in Fig.~\ref{f2}(b).
Note that the profiles in the high-angle detector banks indicate no ferromagnetic component for the sample with $x$ of 0.986 since the intensity of the (002) peak shows no difference between 3 K and 180 K. The details will be published elsewhere.

Figure~\ref{f2}(c) shows the temperature dependence of the SANS intensity, derived by subtracting the \SI{180}{\kelvin} data as background.
Both samples showed a gradual increase in the magnetic intensity below \SI{150}{\kelvin} with cooling $T$.
The intensity increase for $x$ of 1.007 indicates the presence of ferromagnetism.
Figure~\ref{f2}(d) shows the temperature dependence of magnetic satellite peak position $Q_{\rm mag}$.
Note that the profiles were well fitted with a Gaussian function for the $Q_{\rm mag}$ and a Porod function for the downward scattering intensity.
The $Q_{\rm mag}$ kept constant in the $T$ regime from 5 to \SI{50}{\kelvin}.
As $T$ rose above 50 K, the $Q_{\rm mag}$ increased and reached a maximum at \SI{100}{\kelvin}.

The present study demonstrates the importance of controlling the nominal Cr amount for the magnetic ordering in Cr$_{x}$Ta$_{3}$S$_{6}$.
The SANS measurements revealed the presence of magnetic satellite peaks, which is regarded as evidence of the CHM order, in a particular range of the Cr amount. 

The magnetization curves of the powder samples accompany the sharp increase at the low-$H$ regime, which is in contrast with those of bulk single crystals, where only the downward-convex behavior appears. Such ferromagnetic behavior has been found in microfabricated CrNb$_{3}$S$_{6}$ samples\cite{Mito2018} and should be correlated with the process of the CSL formation. The sample dimensions of the powder samples may induce a similar effect on the magnetization. 

The increase in $Q_{\rm mag}$ with increasing temperature was also observed in CrNb$_{3}$S$_{6}$ thin-film lamellae\cite{Togawa2019}. The origin of this behavior was interpreted based on a two-dimensional melting scenario in a magnetic system\cite{Togawa2019}, which can be derived from the renormalization-group theory\cite{Nosov2017}.
$Q_{\rm mag}$ began to decrease above \SI{100}{\kelvin}.
However, detailed features of this decrease near $T_{\rm c}$ are hindered by large uncertainties in $Q_{\rm mag}$ arising from the weak magnetic signals
relative to the background. Further discussion would require additional measurements such as SANS polarization analysis, which can isolate a purely magnetic contribution.

\begin{acknowledgment}
This work was supported by JSPS KAKENHI (Grant Nos. JP17H02815, JP22H01944, JP23H00091, JP23H00312, JP23H01870, JP24K21737, JP25K00962, JP25K01655, JP26K01406, JP26H02203 and JP26H00678), 
JSPS International Joint Research Program (JRP-LEAD with UKRI) (Grant No. JPJSJRP20241710)
and JST ERATO (Grant No. JPM-JER2503).
The SANS experiments at the MLF of the J-PARC were performed under user programs (Proposal No. 2020C0001).

\end{acknowledgment}

\end{document}